\documentstyle[aps,twocolumn,prl,epsfig]{revtex}

\newcommand{\be}{\begin{equation}}
\newcommand{\ee}{\end{equation}}
\newcommand{\bea}{\begin{eqnarray}}
\newcommand{\eea}{\end{eqnarray}}

\begin{document}

\tighten\draft

\twocolumn[\hsize\textwidth\columnwidth\hsize\csname@twocolumnfalse\endcsname

\title{Inverse Time--Dependent Quantum Mechanics}
\author{J. C. Lemm} 
\address{
Institut f\"ur Theoretische Physik I,
Universit\"at M\"unster, 48149 M\"unster, Germany
}
\date{\today}
\maketitle

\begin{abstract}
Using a Bayesian method
for solving inverse quantum problems,
potentials of quantum systems  
are reconstructed from coordinate measurements in
non--stationary states. 
The approach is based on two basic inputs: 
1.~a likelihood model, 
providing the probabilistic description of the measurement process
as given by the axioms of quantum mechanics, and 
2.~additional {\it a priori} information
implemented in form of stochastic processes over potentials.

\end{abstract}
\pacs{03.65.-w, 02.50.Rj, 02.50.Wp} 
] 

\narrowtext

The first step to be done when applying quantum mechanics
to a real world system is the reconstruction of its Hamiltonian
from observational data.
Such a reconstruction, also known as inverse problem,
constitutes a typical example of empirical learning.
Whereas the determination of potentials 
from spectral and from scattering data 
has been studied in much detail in
inverse spectral and inverse scattering theory 
\cite{iqs:Chadan-Sabatier:1989,iqs:Newton:1989},
this Paper describes 
the reconstruction of potentials 
by measuring particle positions in coordinate space
for finite quantum systems in {\it time--dependent} states.
The presented method can easily be generalized to 
other forms of observational data.

In the last years much effort has been devoted 
to many other practical empirical learning problems,
including, just to name a few,
prediction of financial time series,
medical diagnosis, and image or speech recognition.
This also lead to a variety of new learning algorithms,
which should in principle also be applicable 
to inverse quantum problems.
In particular, this Paper shows how the
{\it Bayesian} framework \cite{iqs:Bernado-Smith:1994}
can be applied to solve
problems of inverse time--dependent quantum mechanics (ITDQ).
The presented method generalizes a recently introduced approach
for stationary quantum systems
\cite{iqs:Lemm:1999b,iqs:Lemm:1999c}.
Compared to stationary inverse problems,
the observational data in time--dependent problems
are more indirectly related to potentials,
making them in general more difficult to solve.

Specifically, we will study 
the following type of observational data:
Preparing a particle 
in an eigenstate of the position operator with coordinates $x_0$
at time $t_0$, we let
this state evolve in time
according to the rules of quantum mechanics
and measure its new position at time $t_1$,
finding a value $x_1$.
Continuing from this measured position $x_1$,
we measure the particle position again at time $t_2$, 
and repeat this procedure
until $n$ data points $x_i$ at times $t_i$ have been collected.
We thus end up with observational data of the form 
$D$ =$\{(x_i,\Delta_i,x_{i-1})|1\le i\le n\}$,
where $x_i$ is the result of the $i$-th coordinate measurement,
$\Delta_i$ = $t_i-t_{i-1}$ the time interval 
between two subsequent measurements
and $x_{i-1}$ the coordinates 
of the previous observation (or preparation) at time $t_{i-1}$.

We will discuss in particular systems with 
time--independent Hamiltonians of the form
$H$ = $T+V$,
consisting of a standard kinetic energy term $T$
and a local potential
$V(x,x^\prime)$ = $\delta(x-x^\prime) v(x)$,
with $x$ denoting the position of the particle.
In that case, the aim is the reconstruction of the function $v(x)$
from observational data $D$. 
(The restriction to local potentials simplifies the numerical calculations.
Nonlocal Hamiltonians can be reconstructed similarly.)

Setting up a Bayesian model
requires the definition of two probabilities:
1.~the probability $p(D|v)$
to measure data $D$ given potential $v$, 
which,
for $D$ considered fixed, 
is also known as the {\it likelihood} of $v$,
and 2.~a prior probability $p(v)$
implementing available {\it a priori} information
concerning the potential to be reconstructed.

Referring to a {\it maximum a posteriori approximation} (MAP)
we understand those potentials $v$ 
to be solutions of the 
reconstruction problem,
which maximize $p(v|D)$, i.e., the 
{\it posterior} probability of $v$ given all available data $D$.
The basic relation is then Bayes' theorem, 
according to which $p(v|D)\propto p(D|v) p(v)$.

One possibility is to choose a parametric ansatz for the potential $v$.
In that case, an additional prior term $p(v)$ is often not included
(so the MAP becomes a maximum likelihood approximation).
In the following, 
we concentrate on nonparametric approaches,
which are less restrictive 
compared to their parametric counterparts.
Their large flexibility, however,
makes it essential to include (nonuniform) priors.
Corresponding nonparametric priors 
are formulated explicitly in terms of the function $v(x)$
\cite{iqs:Wahba:1990}.
Indeed, nonparametric priors are well known from
applications to 
regression \cite{iqs:Williams-Rasmussen:1996},
classification \cite{iqs:Williams-Barber:1998},
general density estimation \cite{iqs:Lemm:1999a},
and stationary inverse quantum problems \cite{iqs:Lemm:1999b,iqs:Lemm:1999c}.
It is the likelihood model, discussed next, which is specific for ITDQ.

According to the axioms of quantum mechanics
the probability that a particle is found at 
position $x_i$ at time $t_i$,
provided the particle has been at $x_{i-1}$ at time $t_{i-1}$,
is given by
\be
p_i = p(x_i|\Delta_i,x_{i-1}, v)
= |\phi_i (x_i)|^2
,
\label{transition}
\ee
where 
\be
\phi_i(x_i) \,=\, <x_i\,|\, \phi_i \!> \,=\, <\!x_i|U_i \,x_{i-1}\!>
,
\label{transition-ampl}
\ee
are matrix elements of the time evolution operator
\be
U_i = e^{-i\Delta_i H}
, 
\ee
setting $\hbar$ = 1.
The transition amplitudes (\ref{transition-ampl})
can be calculated by inserting 
orthonormalized eigenstates $\psi_\alpha$ of $H$,
with energies $E_\alpha$,
\be
\phi_i(x_i) 
= \sum_\alpha e^{-i\Delta_i E_\alpha} \psi_\alpha(x_i)\psi_\alpha^*(x_{i-1})
.
\label{energy-phi}
\ee
Clearly, it is straightforward to modify
(\ref{transition}) for measuring
observables different from the particle position.
It is also interesting to note that 
the transition probabilities (\ref{transition})
define a Markoff process with
$W_i(x\rightarrow x^\prime)$
= $p(x^\prime |\Delta_i ,x,v)$.
For real eigenfunctions $\psi_\alpha(x)$, 
i.e., for a real Hamiltonian
with real boundary conditions,
they obey the relation 
$W_i(x\rightarrow x^\prime)$
= $W_i(x^\prime \rightarrow x)$.
It follows that the detailed balance condition,
$p_{\rm stat}(x) W_i(x\rightarrow x^\prime)$
=
$p_{\rm stat}(x^\prime) W_i(x^\prime\rightarrow x)$,
is fulfilled for a uniform 
$p_{\rm stat}(x)$,
which therefore represents the stationary state
of the Markoff process of repeated position 
measurements.

Having defined the likelihood model of ITDQ,
in the next step a prior for $v$ has to be chosen.
A convenient 
nonparametric prior $p(v)$ is a Gaussian 
\begin{equation}
p_G(v) = 
\left(\det \frac{{\bf K}_0}{2\pi}\right)^\frac{1}{2}
e^{-\frac{1}{2} < v-v_0 | {\bf K}_0 | v-v_0 >}
,
\label{gaussprior}
\end{equation}
with (real symmetric, positive semi--definite) inverse covariance
${\bf K}_0$, acting in the space of potentials, 
and mean $v_0(x)$, which can be
considered as a reference potential
for $v$.
Typical examples are smoothness constraints on $v$ 
which correspond to 
choosing differential operators for ${\bf K}_0$.
Reference potentials can be made more flexible
by allowing parameterized families $v_0(x;\theta)$.
Within the context of Bayesian statistics 
such additional parameters $\theta$ are
known as hyperparameters.
In MAP approximation the optimal hyperparameters 
are determined 
by maximizing the posterior (\ref{posterior})
simultaneously with respect to $\theta$ and $v(x)$
\cite{iqs:MacKay:1994d}.
A simplified procedure consists in using
a parametric approximation $v(\theta)$
which maximizes the likelihood $\prod_i p_i(\theta)$
as reference potential $v_0$ for the nonparametric reconstruction $v(x)$
\cite{iqs:Lemm:1999a}.

If available, it is useful to include 
some information about the ground state energy $E_0(v)$,
which helps to determine the depth of the potential.
This can, for example, 
be a noisy measurement of the ground state energy
which, assuming Gaussian noise, 
is implemented by 
\be
p_E\propto e^{-\frac{\mu}{2}\left(E_0(v)-\kappa\right)^2}
.
\label{energy-factor}
\ee

Combining 
(\ref{gaussprior})
and
(\ref{energy-factor})
with (\ref{transition}) for
$n$ repeated coordinate measurements
starting from an initial position $x_0$,
we obtain for the posterior (\ref{posterior}),
\be
p(v|D) 
\propto p_G(v)\, p_E(v)\prod_{i=1}^n p_i
\label{posterior}
.
\ee
To calculate the MAP solution
$v^*$ =${\rm argmax}_v p(v|D)$
we set the functional derivative 
of the posterior (\ref{posterior}),
or technically more convenient of its logarithm, 
with respect to $v$, denoted $\delta_v$,
to zero. This yields,
\be
0 =
\delta_{v} \ln p(v|D)
=
\delta_{v} \ln p_G(v)
+
\delta_{v} \ln p_E(v)
+
\sum_i \delta_{v} \ln p_i  
,
\label{stat-Eq}
\ee
with
\begin{eqnarray}
\delta_{v} \ln p_G(v) 
&=& -{\bf K}_0\,(v-v_0)
,
\label{prior-dev}
\\
\delta_{v} \ln p_E(v) 
&=& -\mu\big(E_0(v)-\kappa\big)\,\delta_{v} E_0(v) 
,
\label{ener-dev}
\\
\delta_{v} \ln p_i 
&=& 2 {\rm Re} [\phi_i^{-1}(x_i) \, \delta_{v} \phi_i(x_i)] 
\label{data-dev}
.
\end{eqnarray}
The functional derivative 
$\delta_v\phi_i$
can, according to Eq.~(\ref{energy-phi}), be obtained from 
$\delta_v \psi_\alpha$.
The still required
$\delta_v \psi_\alpha$
and $\delta_v E_\alpha$
can then be found by calculating the functional derivative
of the eigenvalue equation 
$H\psi_\alpha$ = $E_\alpha\psi_\alpha$.
Using
\be
\delta_{v(x)} V(x^\prime,x^{\prime\prime}) 
= \delta(x-x^\prime)\delta(x^\prime-x^{\prime\prime})
,
\ee
$\delta_{v(x)}$ denoting
the $x$ component of functional derivative $\delta_{v}$,
we find,
\bea
\delta_{v(x)} E_\alpha 
&=& <\!\!\psi_\alpha |\, \delta_{v(x)} H \,| \psi_\alpha\!\!>
=|\psi_\alpha(x)|^2
,
\label{deltaE}
\\
\delta_{v(x)} \psi_\alpha(x^{\prime})
&=& 
\sum_{\gamma\ne \alpha} \frac{1}{E_\alpha-E_\gamma}\, 
\psi_\gamma(x^{\prime})\psi^*_\gamma(x) \psi_\alpha (x)
.
\eea
Collecting the results, gives
\bea
&&
\delta_{v(x)} \phi_i(x_i)
=
\delta_{v(x)} <\!x_i|U_i\,x_{i-1}\!> \,
=
\nonumber\\&&
\sum_\alpha e^{-i\Delta_i E_\alpha} 
\Big[
\left(-i\Delta_i |\psi_\alpha(x)|^2)\right)
            \psi_\alpha(x_i)\psi_\alpha^*(x_{i-1})
\nonumber\\&&
+
\sum_{\gamma\ne \alpha} \frac{1}{E_\alpha-E_\gamma} 
\psi_\gamma(x_i)\psi^*_\gamma(x) \psi_\alpha (x)
\psi_\alpha^*(x_{i-1})
\nonumber\\&&
+
\sum_{\gamma\ne \alpha} \frac{1}{E_\alpha-E_\gamma} 
\psi_\gamma^*(x_{i-1})\psi_\gamma(x) \psi_\alpha^* (x)
\psi_\alpha(x_i)
\Big]
.
\label{delta-phi}
\eea
Inserting 
Eq.~(\ref{deltaE}) for $\alpha$ = 0 in 
Eq.~(\ref{ener-dev})
and Eq.~(\ref{delta-phi})
in Eq.~(\ref{data-dev})
a MAP solution for the potential $v$ can be found
by iterating the stationarity equation (\ref{stat-Eq})
numerically on a lattice.
Clearly, such a straightforward discretization
can only be expected to work for a low--dimensional $x$ variable.
Higher dimensional systems usually 
require additional approximations \cite{iqs:Lemm:1999c}.

\begin{figure}
\begin{center}
\epsfig{file=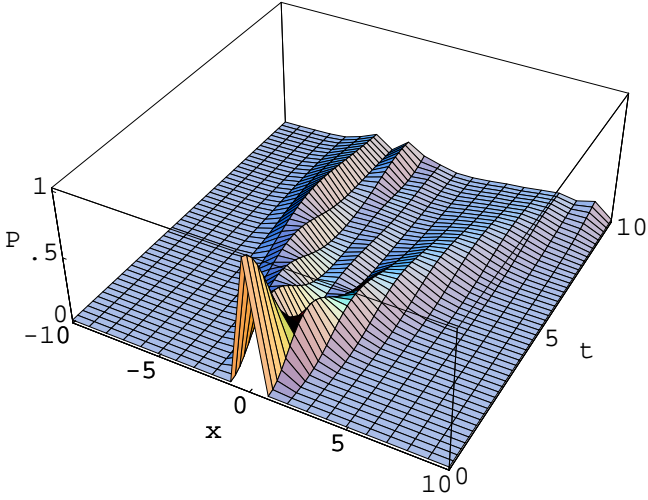, width= 86mm}
\end{center}
\caption{Time evolution of an unobserved particle
with mass $m$ = 1 started at time $t_0$ = 0 from $x_0$ = 0 
in the potential (\ref{true-potential}).
Shown is the transition probability $p(x|\Delta t = t,x_0=1,v_{\rm true})$.}
\label{td-TP}
\end{figure}

\begin{figure}
\begin{center}
\epsfig{file=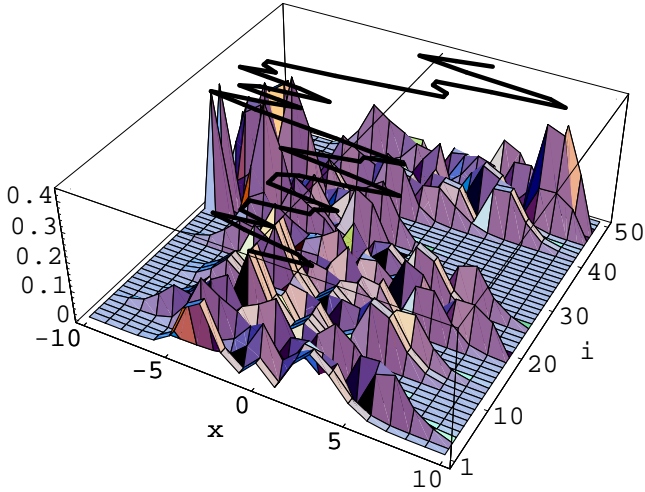, width= 86mm}
\end{center}
\caption{Time evolution of an observed particle with mass $m$ = 1
in the potential (\ref{true-potential}).
The figure shows for each data point $i$
the probability $p(x|\Delta_i=5,x_{i-1},v_{\rm true})$,
starting from $x_0$ = 0.
(Hence, the probability at $i$ =1 corresponds to that shown in 
Fig.~\ref{td-TP} at $t$ = 5.)
The actual data points $x_i$ have been sampled from that
probabilities and form the observed path  
shown on top as a thick line.
}
\label{td-path}
\end{figure}

As the next step,
we want to check the numerical feasibility of a nonparametric reconstruction
of the potential $v$
for a one--dimensional quantum system.
For that purpose, we choose
a system with the true potential
\be
v_{\rm true}(x)
= \frac{c_1}{\sqrt{2\pi\sigma}}e^{\frac{(x-c_2)^2}{2\sigma^2}}
,
\label{true-potential}
\ee
where
$c_1$ = $-10$, 
$c_2$ = $-2$, 
and $\sigma$ = 2.
An example of the time evolution of an unobserved particle 
in the potential $v_{\rm true}$ is shown in Fig.~\ref{td-TP}.
As input for the reconstruction algorithm 
50 data points $x_i$ are sampled from the corresponding true likelihoods
$p(x|\Delta_i,x_{i-1},v_{\rm true})$.
A corresponding path of an observed particle 
is shown in  Fig.~\ref{td-path}.

\begin{figure}
\begin{center}
\epsfig{file=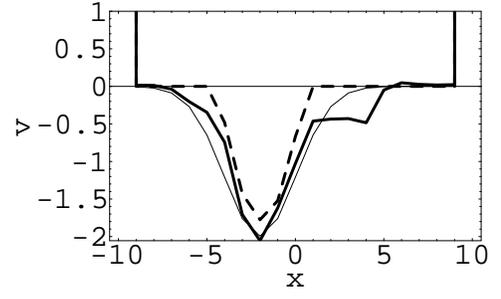, width= 65mm}
\end{center}
\caption{Numerical reconstruction of a potential 
from 50 coordinate measurements
(see Fig.~\ref{td-path}).
Shown are
the true potential $v_{\rm true}$ (thin line),
the best parametric approximation 
used as reference potential $v_0$ (dashed line),
and the reconstructed potential $v_{\rm ITDQ}$ (thick line).
Parameters: $m$ = 1,
$\Delta_i$ = 5,
$v_{\rm true}$ of Eq.~(\ref{true-potential}),
$v_0$ of the form (\ref{paramV}),
Gaussian prior with 
${\bf K}_0$ as in (\ref{inv-cov}) 
with $\lambda$ = 0.1 and $\sigma_{0}$ = 3,
$p_E$ (\ref{energy-factor})
with $\mu$ = 10 and $\kappa$ = $E_0(v_{\rm true})$, 
periodic boundary conditions for $\psi_\alpha$,
fixed boundary values $v(-10)$ = $v(10)$ = $10^5$
for $v_{\rm ITDQ}$, $v_0$ and $v_{\rm true}$,
calculated on a lattice with 21 points. 
Errors:
$\epsilon_D(v_{\rm ITDQ})$ = 99.1,
$\epsilon_D(v_{\rm true})$ = 104.4,
$\epsilon_g(v_{\rm ITDQ})$ = 1.891,
$\epsilon_g(v_{\rm true})$ = 1.818.
}
%
%
%
%
%
%
%
\label{td-fig1a}
\end{figure}

Besides a noisy energy measurement of the form (\ref{energy-factor})
we include a Gaussian prior (\ref{gaussprior})
with a smoothness related inverse covariance
\be
{\bf K}_0(x,x^\prime)
= 
\delta(x-x^\prime)\lambda\sum_{k=0}^3 (-1)^k \frac{\sigma_{0}^{2m}}{k!2^k}
\left(\frac{\partial^2}{\partial x^2}\right)^k
.
\label{inv-cov}
\ee
To obtain an adapted reference potential $v_0$ for the Gaussian prior, 
a parameterized potential of the form
\be
v_0(a,b,c)
=
{\rm min}[0,a(x-b)^2+c]
\label{paramV}
,
\ee
is optimized with respect to $a$, $b$, $c$ 
by maximizing the ``extended likelihood''
$\sum_i\ln p_i(v_0)+\ln p_E(v_0)$.
Finally,
the stationarity equation (\ref{stat-Eq})
is solved by iterating according to
\bea
&v^{(r+1)} 
= v^{(r)} + \eta \Big[ v_0-v^{(r)}+
&\\&
{\bf K}_0^{-1}\big\{
2\sum_i^n {\rm Re} [\delta_{v} \ln\phi_i(x_i)]
+ \delta_{v}(\ln p_G+\ln p_E)\big\}\Big]
.&
\nonumber
\eea
The resulting nonparametric ITDQ solution $v_{\rm ITDQ}$ 
(see Fig.~\ref{td-fig1a}),
is a reasonable reconstruction of $v_{\rm true}$,
and clearly better than the best parametric approximation $v_0(a,b,c)$.
It is only the flat area near the right border
where, due to missing and unrepresentative data, 
the reconstruction differs significantly from the true potential.

\begin{figure}
\begin{center}
\epsfig{file=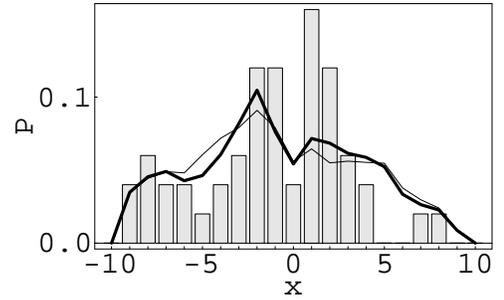, width= 65mm}
\end{center}
\caption{Sum of empirical transition probabilities 
$p_{\rm emp}$ 
(bars),
the corresponding true 
$p_{\rm true}$ 
(thick line)
for $v_{\rm true}$,
and the reconstructed
$p_{\rm ITDQ}$ 
(thin line).}
\label{td-fig1b}
\end{figure}

\begin{figure}
\begin{center}
\epsfig{file=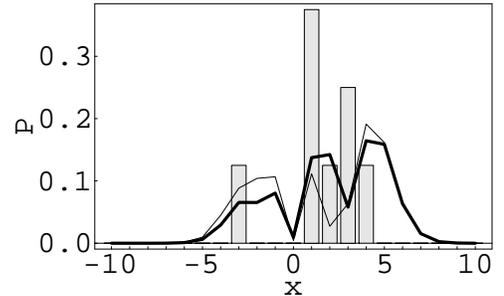, width= 65mm}
\end{center}
\caption{Same functions as in Fig.~\ref{td-fig1b},
but restricted to measurements of a particle
which has been at position
$x_{i-1}$ = 1 at the time of the previous measurement.
}
\label{td-fig1c}
\end{figure}

Fig.~\ref{td-fig1b} 
compares the sum over empirical transition probabilities
$\frac{1}{n}\sum_{i=1}^n\delta (x-x_i)$ as 
derived from the observational data $D$
with the corresponding true 
$p_{\rm true}$ = $\frac{1}{n}\sum_{i=1}^n p(x|\Delta_i,x_{i-1},v_{\rm true})$ 
and reconstructed 
$p_{\rm ITDQ}$ = $\frac{1}{n}\sum_{i=1}^n p(x|\Delta_i,x_{i-1},v_{\rm ITDQ})$.
Due to the summation over data points with different $x_{i-1}$,
the quantities shown in Fig.~\ref{td-fig1b} 
do not present the complete information which is available to the algorithm.
Hence, Fig.~\ref{td-fig1c} depicts the corresponding quantities
for a fixed $x_{i-1}$.
In particular, Fig.~\ref{td-fig1c} compares
the reconstructed transition probability (\ref{transition})
with the corresponding empirical and true transition probabilities
for a particle having been at time $t_{i-1}$ at position $x_{i-1}$ =1.
The ITDQ algorithm returns an 
approximation for all such transition probabilities.

Figs.~\ref{td-fig1b} and \ref{td-fig1c} show,
that the reconstructed $v_{\rm ITDQ}$
tends to produce a better approximation of the 
empirical probabilities
than the true potential $v_{\rm true}$.
Indeed, 
the error on the data or negative log--likelihood,
$\epsilon_D(v)$ = $-\sum_i\ln p_i(v)$,
being a canonical error measure in density estimation,
is smaller for $v_{\rm ITDQ}$ than for $v_{\rm true}$.
A smaller $\lambda$, 
i.e., a lower influence of the prior,
produces a still smaller error
$\epsilon_D(v_{\rm ITDQ})$. 
At the same time, however, 
the reconstructed potential becomes more wiggly for smaller $\lambda$,
being the symptom of the well known effect of ``overfitting''.
The (true) generalization error 
$\epsilon_g(v)$ = 
$-\int\!dx\,dx^\prime p(x) 
p(x^\prime|x,v_{\rm true})
\ln p(x^\prime|x,v)$
[with uniform $p(x)$], on the other hand,
can never be smaller for the reconstructed $v_{\rm ITDQ}$ 
than for $v_{\rm true}$.
As it is typical for most empirical learning problems,
the generalization error $\epsilon_g(v_{\rm ITDQ})$ 
shows a minimum as function of $\lambda$.
It is this minimum which gives the optimal value for $\lambda$.
Know\-ledge of the true model
allows in our case to calculate the generalization error exactly.
If, as usual, the true model is not known,
classical cross--validation \cite{iqs:Wahba:1990}
and bootstrap \cite{iqs:Efron-Tibshirani:1993} techniques
can be used
to approximate the generalization error as function of $\lambda$ empirically. 

Alternatively to optimizing $\lambda$ 
or other hyperparameters
one can integrate over them \cite{iqs:MacKay:1994d}.
Similarly, studying the feasibility 
of a Bayesian Monte Carlo approach,
contrasting the MAP approach of this paper,
would certainly be interesting.

In summary,
this Paper has presented a method 
to solve inverse problems for time--dependent quantum systems.
The approach, based on a Bayesian framework, 
is able to handle quite general types of observational data. 
Numerical calculations proved to be feasible 
for a one dimensional model.


\begin{thebibliography}{10}

\bibitem{iqs:Chadan-Sabatier:1989}
K. Chadan and P.~C. Sabatier, {\em Inverse Problems in Quantum Scattering
  Theory} (Springer Verlag, New York, 1989).

\bibitem{iqs:Newton:1989}
R.~G. Newton, {\em Inverse Schr\"odinger Scattering in Three Dimensions}
  (Springer Verlag, Berlin, 1989).

\bibitem{iqs:Bernado-Smith:1994}
J.~M. Bernado and A.~F. Smith, {\em Bayesian Theory} (Wiley, New York, 1994).

\bibitem{iqs:Lemm:1999b}
J.~C. Lemm, J. Uhlig, and A. Weiguny, Technical Report No.~MS-TP1-99-6, Univ.\
  of M\"unster, arXiv:cond-mat/9907013 (accepted for
  publication by {\it Phys. Rev. Lett.}

\bibitem{iqs:Lemm:1999c}
J.~C. Lemm and J. Uhlig, Technical Report No.~MS-TP1-99-10, Univ.\ of
  M\"unster, arXiv:nucl-th/9908056.

\bibitem{iqs:Wahba:1990}
G. Wahba, {\em Spline Models for Observational Data} (SIAM, Philadelphia,
  1990).

\bibitem{iqs:Williams-Rasmussen:1996}
C.~K.~I. Williams and C.~E. Rasmussen,  in {\em NIPS 8}, edited by D.~S.
  Touretzky, M.~C. Mozer, and M.~E. Hasselmo (The {MIT} Press, Cambridge, MA,
  1996), pp.\ 514--520.

\bibitem{iqs:Williams-Barber:1998}
C.~K.~I. Williams and D. Barber, IEEE Trans. on Pattern Analysis and Machine
  Intelligence {\bf 20},  1342  (1998).

\bibitem{iqs:Lemm:1999a}
J.~C. Lemm, Technical Report No.~MS-TP1-99-1, Univ.\ of M\"unster, 
arXiv:physics/9912005.

\bibitem{iqs:MacKay:1994d}
D.~J.~C. MacKay,  in {\em Maximum Entropy and Bayesian Methods, Santa Barbara
  1993.}, edited by G. Heidbreder (Kluwer, Dordrecht, 1994).

\bibitem{iqs:Efron-Tibshirani:1993}
B. Efron and R.~J. Tibshirani, {\em An Introduction to the Bootstrap} (Chapman
  \& Hall, New York, 1993).

\end{thebibliography}

\vspace{3cm}

\end{document}